# Matrix representation of the resolvent operator in square-integrable basis and physical application

A. D. Alhaidari

*Saudi Center for Theoretical Physics, P.O. Box 32741, Jeddah 21438, Saudi Arabia*

**Abstract:** We obtain simple formulas for the matrix elements of the resolvent operator (the Green's function) in any finite set of square integrable basis. These formulas are suitable for numerical computations whether the basis elements are orthogonal or not. A byproduct of our findings is an expression for the normalized eigenvectors of a matrix in terms of its eigenvalues. We give a physical application as an illustration of how useful these results can be.

**Keywords**: resolvent operator, basis sets, eigenvalues and eigenvectors, Green's function.

## 1. Introduction

This work is a detailed and comprehensive but brief treatment of a relevant object frequently encountered in various branches of science and engineering. Such a complete technical treatment of this object (the resolvent operator) is not easily found as single and short document in the literature making this study very useful for researchers and practitioners. The reader is assumed to be familiar with only the basics of linear algebra and matrix manipulations.

In many branches of science and engineering, one frequently comes across various operators that represent certain processes and/or actions. An important object associated with such an operator is the resolvent that captures its spectral property. If we designate the operator as $\mathcal{L}$, then its resolvent is written formally as the inverse operator $\left(\mathcal{L}-z\mathcal{I}\right)^{-1}$ where $\mathcal{I}$ is the identity and $z$ is generally a complex number. It should be obvious that there exist a set of numbers in the complex $z$-plane on which the resolvent operator cannot be realized/defined. Such a set of values, $\{\xi\}$, are those at which the action of the operator on an element of its domain $|\phi\rangle$ becomes $\mathcal{L}:|\phi\rangle \mapsto \xi|\phi\rangle$. The set $\{\xi\}$ is called the "spectral set" of $\mathcal{L}$, which can be discrete or continuous or a combination of both. Moreover, the discrete set can either be finite or countably infinite whereas the continuous set generally consists of several disconnected but continuous regions of space in the complex plane. These regions might be curves or areas. However, the intersection of all of these spectral sets is null. Nonetheless, if the set $\{\xi\}$ is "too large" then one may not be able to find a region in the complex plane in which the resolvent could be realized. However, in this work we will assume that such scenario does not occur and we'll be able to define a resolvent for $\mathcal{L}$ in some region of the complex plane and thus gain knowledge about its spectral property. A physical example is the Hamiltonian operator of quantum mechanics with $z$ being the energy, $\{\xi\}$ the energy spectrum and $\{|\phi\rangle\}$ the discrete bound states or continuous scattering states. For such systems, the resolvent is usually referred to in the physics literature as the "Green's function".

In the following section, we take $\mathcal{L}$ to be a self-adjoint operator whose action on a discrete set $\{|\phi_n\rangle\}$ is well-defined. Then, we derive the matrix elements of the resolvent on the conjugate

set $\left\{\left|\bar{\phi}_n\right\rangle\right\}$. In Section 3, we obtain an alternative but equivalent representation of the resolvent that is more suitable numerically because it reduces the computational cost. In Section 4, we present a physical application where our findings prove to be very useful in locating resonance energies of a given quantum mechanical system, obtaining its bound states energies, and the density of states. Finally, we conclude in Section 5.

## 2. Preliminaries

Let $\{\psi_n\}_{n=0}^{\infty}$ be a complete set of square-integrable functions in configuration space that supports a Hermitian matrix representation for a self-adjoint operator *H*. The conjugate space is spanned by $\{\bar{\psi}_n\}_{n=0}^{\infty}$, where $\langle\bar{\psi}_n|\psi_m\rangle = \langle\psi_n|\bar{\psi}_m\rangle = \delta_{n,m}$ and $\sum_n|\bar{\psi}_n\rangle\langle\psi_n| = \sum_n|\psi_n\rangle\langle\bar{\psi}_n| = 1$. The first of these two relations is called the "orthogonality" relation and the second the "completeness" statement. The resolvent, which is also known as the Green's function $G(z)$, is formally defined by $G(z)\left(H-z\right)=1$ where *z* is a real number. Since the matrix elements of *H* are given in the basis $\{\psi_n\}_{n=0}^{\infty}$ as $H_{n,m} = \langle\psi_n|H|\psi_m\rangle$, then those of $G(z)$ are given in the conjugate basis as follows

$$G_{n,m}(z) = \langle\bar{\psi}_n|(H-z)^{-1}|\bar{\psi}_m\rangle \tag{1}$$

Now, numerical manipulation of the resolvent, which involves taking the inverse of operators, is carried out most appropriately and efficiently in an orthonormal basis set $\{\chi_n\}_{n=0}^{\infty}$ (i.e., $\langle\chi_n|\chi_m\rangle = \delta_{n,m}$) in which the representation of those operators is diagonal. That is to say, we start by solving the eigenvalue problem

$$H|\chi_n\rangle = \varepsilon_n|\chi_n\rangle \tag{2}$$

From now on, we work in a finite subspace of dimension *N* and obtain the finite *N*-dimensional representation of the Green's function $G_{n,m}^N(z)$ where $G_{n,m}(z) = \lim_{N\to\infty} G_{n,m}^N(z)$. Since the matrix representations of the relevant operators are in the basis $\{\psi_n\}$ rather than $\{\chi_n\}$, then we can rewrite Eq. (2) as follows

$$\sum_{k=0}^{N-1}\langle\psi_m|H|\psi_k\rangle\langle\bar{\psi}_k|\chi_n\rangle = \varepsilon_n\sum_{k=0}^{N-1}\langle\psi_m|\psi_k\rangle\langle\bar{\psi}_k|\chi_n\rangle \quad ;n,m=0,1,..,N-1 \tag{3}$$

where we have used the completeness property of the basis in the finite *N* dimensional subspace,

$$\sum_k|\psi_k\rangle\langle\bar{\psi}_k| = \sum_k|\bar{\psi}_k\rangle\langle\psi_k| = I\,, \tag{4}$$

and *I* is the $N\times N$ unit matrix. In matrix notation, Eq. (3) reads

$$\sum_{k=0}^{N-1}H_{m,k}\Gamma_{k,n} = \varepsilon_n\sum_{k=0}^{N-1}\Omega_{m,k}\Gamma_{k,n} \quad ;n,m=0,1,..,N-1 \tag{5}$$

where $\Gamma_{k,n} := \langle \bar{\psi}_k | \chi_n \rangle$ and $\Omega$ is the basis overlap matrix whose elements are $\Omega_{n,m} = \langle \psi_n | \psi_m \rangle$. Thus, $\left\{ \Gamma_{k,n} \right\}_{k=0}^{N-1}$ is the generalized eigenvector associated with the generalized eigenvalue $\varepsilon_n$. This is so because Eq. (5) could be written in matrix notation as the following generalized eigenvalue equation in the $\{\psi_n\}$ basis

$$H|\Gamma_n\rangle = \varepsilon_n \Omega |\Gamma_n\rangle. \tag{6}$$

Moreover, Eq. (5) reads $(H\Gamma)_{m,n} = \varepsilon_n (\Omega\Gamma)_{m,n}$ which when multiplied from left by $\Gamma^{\mathrm{T}}$, where $\Gamma^{\mathrm{T}}{}_{n,m} = \langle \chi_n | \bar{\psi}_m \rangle$, becomes

$$(\Gamma^{\mathrm{T}} H \Gamma)_{m,n} = \varepsilon_n (\Gamma^{\mathrm{T}} \Omega \Gamma)_{m,n} \qquad ; n,m = 0,1,..,N-1 \tag{7}$$

Being the generalized eigenvectors associated with the matrices $H$ and $\Omega$, the normalized eigenvector matrix $\Gamma$ simultaneously diagonalizes $H$ and $\Omega$. That is,

$$(\Gamma^{\mathrm{T}} H \Gamma)_{n,m} = \eta_n \delta_{n,m} \qquad \text{and} \qquad (\Gamma^{\mathrm{T}} \Omega \Gamma)_{n,m} = \sigma_n \delta_{n,m} \tag{8}$$

Henceforth, we deduce that $\varepsilon_n = \eta_n / \sigma_n$ and equation (1) could be written as

$$\begin{aligned} G_{n,m}^N(z) &= \sum_{i,j,k,l=0}^{N-1} \langle \bar{\psi}_n | \chi_i \rangle \langle \chi_i | \bar{\psi}_k \rangle \langle \psi_k | (H - z)^{-1} | \psi_l \rangle \langle \bar{\psi}_l | \chi_j \rangle \langle \chi_j | \bar{\psi}_m \rangle \\ &= \sum_{i,j,k,l=0}^{N-1} \Gamma_{n,i} \left\{ \Gamma_{i,k}^{\mathrm{T}} \left[ (H - z\Omega)^{-1} \right]_{k,l} \Gamma_{l,j} \right\} \Gamma_{j,m}^{\mathrm{T}} \end{aligned} \tag{9}$$

Now,

$$\sum_{k,l=0}^{N-1} \Gamma_{i,k}^{\mathrm{T}} \left[ (H - z\Omega)^{-1} \right]_{k,l} \Gamma_{l,j} = \frac{\delta_{i,j}}{\eta_i - z\sigma_i} = \frac{1}{\sigma_i} \frac{\delta_{i,j}}{\varepsilon_i - z} \tag{10}$$

Therefore, we can write (9) as follows

$$\boxed{G_{n,m}^N(z) = \sum_{i=0}^{N-1} \frac{\Gamma_{n,i}\Gamma_{m,i}}{\eta_i - z\sigma_i} = \sum_{i=0}^{N-1} \frac{1}{\sigma_i} \frac{\Gamma_{n,i}\Gamma_{m,i}}{\varepsilon_i - z}}. \tag{11}$$

For orthonormal basis (when $\bar{\psi}_n = \psi_n$) the overlap matrix $\Omega$ is just the unit matrix $I$, hence $\sigma_i = 1$, $\eta_i = \varepsilon_i$ . In this orthonormal basis, we can write

$$\boxed{G_{n,m}^N(z) = \sum_{i=0}^{N-1} \frac{\Gamma_{n,i}\Gamma_{m,i}}{\varepsilon_i - z}}. \quad \text{(in orthonormal basis)} \tag{12}$$

The terms "orthonormal" and "non-orthonormal" in the language of bases correspond (in the language of matrix equations) to the terms "eigenvalue equation" and "generalized eigenvalue equation", respectively.

## 3. Alternative representation

To avoid the sumptuous calculation of eigenvectors of matrices and reduce the computational cost, it is preferred that we work with eigenvalues of matrices rather than their eigenvectors. Here we give an alternative formula to (11) and (12) where only eigenvalues are involved. Let $H^{(n,m)}$ and $\Omega^{(n,m)}$ be the $(N-1)\times(N-1)$ submatrices of $H$ and $\Omega$ obtained by deleting row $n$ and column $m$, respectively. The eigenvalue equation and generalized eigenvalue equation in the truncated space, which parallel equations (2) and (6), are

$$H^{(n,m)}\left|\tilde{\chi}_k\right\rangle = \varepsilon_k^{(n,m)}\left|\tilde{\chi}_k\right\rangle, \tag{13}$$

$$H^{(n,m)}\left|\tilde{\Gamma}_k\right\rangle = \varepsilon_k^{(n,m)}\Omega^{(n,m)}\left|\tilde{\Gamma}_k\right\rangle, \tag{14}$$

where $k = 0,1,..,N-2$ and $\tilde{\Gamma}_{i,j} := \left\langle \overline{\psi}_i \middle| \tilde{\chi}_j \right\rangle$. Moreover, the matrix of the normalized eigen-vectors, $\tilde{\Gamma}$, simultaneously diagonalizes $H^{(n,m)}$ and $\Omega^{(n,m)}$. That is,

$$\left[\tilde{\Gamma}^{\mathrm{T}} H^{(n,m)}\tilde{\Gamma}\right]_{i,j} = \eta_i^{(n,m)}\delta_{i,j} \qquad \text{and} \qquad \left[\tilde{\Gamma}^{\mathrm{T}}\Omega^{(n,m)}\tilde{\Gamma}\right]_{i,j} = \sigma_i^{(n,m)}\delta_{i,j}\,, \tag{15}$$

with $\varepsilon_k^{(n,m)} = \eta_k^{(n,m)} \big/ \sigma_k^{(n,m)}$. An identity for an $N\times N$ non-singular matrix $C$, which is very useful in such calculation reads

$$(C^{-1})_{n,m} = (-1)^{n+m}\frac{\left|C^{(n,m)}\right|}{|C|} = (-1)^{n+m}\frac{\prod_{i=0}^{N-2} c_i^{(n,m)}}{\prod_{j=0}^{N-1} c_j}\,, \tag{16}$$

where we have also used the determinant identity for $C$ that reads $|C| = \left|\Lambda^{\mathrm{T}} C\,\Lambda\right| = \prod_{n=0}^{N-1} c_n$ with $\{c_n\}_{n=0}^{N-1}$ being the set of its eigenvalues and $\{\Lambda_{m,n}\}_{n,m=0}^{N-1}$ the corresponding normalized eigenvectors. Then, it is easy to show that the following is an alternative but equivalent form for $G_{n,m}^N(z)$ in a non-orthogonal basis

$$\boxed{\begin{aligned} G_{n,m}^N(z) &= (-1)^{n+m}\frac{\left|\Omega^{(n,m)}\right|}{|\Omega|}\left[\prod_{i=0}^{N-2}\left(\varepsilon_i^{(n,m)} - z\right)\Big/\prod_{j=0}^{N-1}\left(\varepsilon_j - z\right)\right] \\ &= (-1)^{n+m}\frac{\prod_{i=0}^{N-2}\tau_i^{(n,m)}}{\prod_{j=0}^{N-1}\tau_j}\left[\prod_{i=0}^{N-2}\left(\varepsilon_i^{(n,m)} - z\right)\Big/\prod_{j=0}^{N-1}\left(\varepsilon_j - z\right)\right] \end{aligned}} \quad \text{(in non-orthogonal basis)} \tag{17}$$

where $\{\tau_n\}_{n=0}^{N-1}$ and $\{\tau_k^{(n,m)}\}_{k=0}^{N-2}$ are the eigenvalues of the overlap matrices $\Omega$ and $\Omega^{(n,m)}$, respectively. For an orthonormal basis, $\Omega^{(n,m)} = I^{(n,m)}$ and (17) is not valid if $n \neq m$ because then $I^{(n,m)}$ is singular. However, a valid formula for orthonormal bases that replaces Eq. (17) is obtained using the identity (16) and reads

$$G_{n,m}^{N}(z)=\left(-1\right)^{n+m}\frac{\prod_{i=0}^{N-2}\omega_i^{(n,m)}(z)}{\prod_{j=0}^{N-1}\varepsilon_j-z}, \qquad \text{(in orthonormal basis)} \qquad (18)$$

where $\left\{\omega_i^{(n,m)}(z)\right\}_{i=0}^{N-2}$ are the eigenvalues of the matrix function $H^{(n,m)}-z\,I^{(n,m)}$. For $n=m$, Eq. (18) becomes

$$\boxed{G_{n,n}^{N}(z)=\frac{\prod_{i=0}^{N-2}\varepsilon_i^{(n,n)}-z}{\prod_{j=0}^{N-1}\varepsilon_j-z}}. \qquad \text{(in orthonormal basis)} \qquad (19)$$

For $n\neq m$, an alternative formula to (18) is preferred to avoid the functional dependence of the eigenvalue $\omega_i^{(n,m)}(z)$ that greatly increases the computational cost. Due to the fact that $I^{(n,m)}$ is singular for $n\neq m$, a formula similar to (19) is not possible. However, because of the product $\prod_{j=0}^{N-1}\varepsilon_j-z$ in the denominator of $G_{n,m}^{N}(z)$ as shown in (18), we can recast $G_{n,m}^{N}(z)$ as the sum $G_{n,m}^{N}(z)=\sum_{i=0}^{N-1}\frac{A_i^{(n,m)}}{\varepsilon_i-z}$, where

$$A_i^{(n,m)}=(-1)^{n+m}\frac{\left|H^{(n,m)}-\varepsilon_i\,I^{(n,m)}\right|}{\prod_{\substack{j=0\\ j\neq i}}^{N-1}\varepsilon_j-\varepsilon_i}. \qquad (20)$$

Using the identity noted above, which equates the determinant of a square matrix to the product of its eigenvalues, we can finally write (17) in an orthonormal basis as follows

$$\boxed{G_{n,m}^{N}(z)=(-1)^{n+m}\sum_{j=0}^{N-1}\frac{\prod_{i=0}^{N-2}\varepsilon_{i,j}^{(n,m)}}{\left(\varepsilon_j-z\right)\prod_{k\neq j}^{N-1}(\varepsilon_k-\varepsilon_j)}}, \qquad \text{(in orthonormal basis)} \quad (21)$$

where $\left\{\varepsilon_{i,j}^{(n,m)}\right\}_{i=0}^{N-2}$ are the $N-1$ eigenvalues of the matrix $H^{(n,m)}-\varepsilon_j\,I^{(n,m)}$. That is, $\varepsilon_{i,j}^{(n,m)}$ = $\omega_i^{(n,m)}(\varepsilon_j)$.

A by-product of formulas (12) and (18) is an interesting recipe for calculating the square of the elements of the normalized eigenvectors of a square Hermitian matrix in terms of its set of eigenvalues as follows

$$\Gamma_{n,k}\Gamma_{m,k}=(-1)^{n+m}\frac{\prod_{i=0}^{N-2}\varepsilon_{i,k}^{(n,m)}}{\prod_{\substack{j=0\\ j\neq k}}^{N-1}\varepsilon_j-\varepsilon_k}, \qquad \text{(in orthonormal basis)} \qquad (22)$$

which is obtained by evaluating (12) and (18) at $z=\varepsilon_k$. However, for $n=m$, we can use (19) instead of (18) to write

$$\boxed{\Gamma_{n,k}^2 = \frac{\prod_{i=0}^{N-2} \varepsilon_i^{(n,n)} - \varepsilon_k}{\prod_{\substack{j=0\\ j\neq k}}^{N-1} \varepsilon_j - \varepsilon_k}}. \quad \text{(in orthonormal basis)} \tag{23}$$

This last formula has been rediscovered recently by D. Mitnik and S. Mitnik [1]. However, we have been utilizing it since the early days of the *J*-matrix method (see, for example, Ref. [2] and citations therein). A version of (22) for non-orthogonal basis could be obtained by combining (11) and (17) where we get

$$\Gamma_{n,k}\Gamma_{m,k} = (-1)^{n+m}\sigma_k \frac{\left|\Omega^{(n,m)}\right|}{|\Omega|} \frac{\prod_{i=0}^{N-2} \varepsilon_i^{(n,m)} - \varepsilon_k}{\prod_{\substack{j=0\\ j\neq k}}^{N-1} \varepsilon_j - \varepsilon_k}. \quad \text{(in non-orthogonal basis)} \tag{24}$$

Moreover, for $n = m$, this gives

$$\boxed{\Gamma_{n,k}^2 = \frac{\sigma_k}{\tau_k} \frac{\prod_{i=0}^{N-2} \tau_i^{(n,n)} \left(\varepsilon_i^{(n,n)} - \varepsilon_k\right)}{\prod_{\substack{j=0\\ j\neq k}}^{N-1} \tau_j \left(\varepsilon_j - \varepsilon_k\right)}}. \quad \text{(in non-orthogonal basis)} \tag{25}$$

# 4. Physical application

As an example, we use the matrix representation of the Green's function derived above to calculate the scattering matrix of a given physical system, obtain its resonance energies, bound state energies, and compute the energy density of states. We consider a system with the Hamiltonian $H = H_0 + V$ where $H_0$ stands for the Hamiltonian operator associated with the Coulomb problem in three dimensions, which has the following realization in the radial coordinate *r*

$$H_0 = \frac{1}{2}\frac{d^2}{dr^2} + \frac{\ell(\ell+1)}{2r^2} + \frac{Z}{r}, \tag{26}$$

where $\ell = 0,1,2,...$ is the angular momentum quantum number, $Z \in \mathbb{R}$ is the electric charge, and we have adopted the atomic units $\hbar = M = e = 1$ in which distances are measured in units of the Bohr radius $4\pi \in_0$. The added potential $V$ is taken as a short-range radial function such that $V(r > R) \approx 0$, where $R$ is some finite range. Now, we take the basis elements as

$$\psi_n(r) = \sqrt{\frac{\lambda\Gamma(n+1)}{\Gamma(n+2\ell+2)}}(\lambda r)^{\ell+1} e^{-\lambda r/2} L_n^{2\ell+1}(\lambda r), \tag{27}$$

where $L_n^{2\ell+1}(x)$ is the Laguerre polynomial and $\lambda$ is a real scale parameter of inverse length dimension. The orthogonality of the Laguerre polynomials gives $\bar{\psi}_n(r) = (\lambda r)^{-1}\psi_n(r)$ and the three-term recursion relation gives

$$\Omega_{n,m} = \langle \psi_n | \psi_m \rangle = 2(n+\ell+1)\delta_{n,m} - \sqrt{n(n+2\ell+1)}\,\delta_{n,m+1} - \sqrt{(n+1)(n+2\ell+2)}\,\delta_{n,m-1}, \quad (28)$$

making $\Omega$ a tridiagonal symmetric matrix. Moreover, using the differential equation of the Laguerre polynomial, its differential property, and recursion relation we obtain the following matrix elements of $H_0$

$$\begin{aligned}(H_0)_{n,m} = \langle \psi_n | H_0 | \psi_m \rangle = \left[\lambda Z + \frac{\lambda^2}{4}(n+\ell+1)\right]\delta_{n,m} \\ + \frac{\lambda^2}{8}\sqrt{n(n+2\ell+1)}\delta_{n,m+1} + \frac{\lambda^2}{8}\sqrt{(n+1)(n+2\ell+2)}\delta_{n,m-1}\end{aligned} \quad (29)$$

Therefore, the matrix elements of *H* in the basis $\{\psi_n\}$ is $H_{n,m} = (H_0)_{n,m} + V_{n,m}$ and since the potential function $V(r)$ is short-range, then its matrix elements $V_{n,m} = \langle \psi_n | V | \psi_m \rangle \approx 0$ if *n* and/or *m* is greater than or equal to some large enough integer *N* that depends on the dimensionless range $\lambda R$. Consequently, the infinite Hermitian matrix *H* will consist of an $N \times N$ block on the top-left corner that represents $H_0 + V$ and an infinite tridiagonal symmetric tail representing $H_0$. That is, the matrix representation of *H* will look like the following

$$H = \begin{pmatrix} \times & \times & \times & \times & \times & \times & \times & & & & & & \\ \times & \times & \times & \times & \times & \times & \times & & & & & & \\ \times & \times & \times & \times & \times & \times & \times & & & & & & \\ \times & \times & \times & \times & \times & \times & \times & & & & & & \\ \times & \times & \times & \times & \times & \times & \times & & & & & & \\ \times & \times & \times & \times & \times & \times & \times & & & & & & \\ \times & \times & \times & \times & \times & \times & \times & b_{N-1} & & & & & \\ & & & & & & b_{N-1} & a_N & b_N & & & & \\ & & & & & & & b_N & a_{N+1} & b_{N+1} & & & \\ & & & & & & & & b_{N+1} & a_{N+2} & \times & & \\ & & & & & & & & & \times & \times & \times & \\ & & & & & & & & & & \times & \times & \times \\ & & & & & & & & & & & \times & \times \end{pmatrix}, \quad (30)$$

where $a_n = \lambda Z + \frac{\lambda^2}{4}(n+\ell+1)$ and $b_n = \frac{\lambda^2}{8}\sqrt{(n+1)(n+2\ell+2)}$.

The solution of the wave equation $H|\Phi\rangle = E|\Phi\rangle$ for all energies (bound states and scattering states) in such a configuration is handled more effectively by using the J-matrix method [3]. In this method, the solution is written as $|\Phi\rangle = \sum_{n=0}^{\infty} f_n |\psi_n\rangle$ and one obtains the coefficients $\{f_n\}$ that depend on $\{E, \ell, Z, \lambda\}$ and the parameters of the potential *V*. The asymptotic coefficients $\{f_n\}_{n=N-1}^{\infty}$ are solutions of the reference wave equation where $V = 0$. They satisfy the following symmetric three-term recursion relation that results from the infinite tail of the matrix wave equation $(H - E\Omega)|\Phi\rangle = 0$ and reads

$$\left[a_n - 2E(n+\ell+1)\right] f_n + \left(1+\tfrac{8E}{\lambda^2}\right) b_{n-1} f_{n-1} + \left(1+\tfrac{8E}{\lambda^2}\right) b_n f_{n+1} = 0\,, \tag{31}$$

for $n = N, N+1, N+2, \ldots$ . This recursion relation has two independent solutions, which we call $c_n(E)$ and $s_n(E)$, and $f_n(E)$ becomes a linear combination of both with energy dependent factors [2,3]. The boundary conditions give

$$f_n(E) = \left[c_n(E) - \mathrm{i}s_n(E)\right] - e^{2\mathrm{i}\delta(E)}\left[c_n(E) + \mathrm{i}s_n(E)\right], \tag{32}$$

for $n = N-1, N, N+1, ..$ and where $\delta(E)$ is the scattering phase shift angle. This phase shift and the $N-1$ coefficients $\{f_n\}_{n=0}^{N-2}$ are determined from the solution of the remaining $N$ equations (after removing the infinite tail from the matrix wave equation) that read

$$\begin{pmatrix} \times & \times & \times & \times & \times & \times & \times \\ \times & \times & \times & \times & \times & \times & \times \\ \times & \times & \times & \times & \times & \times & \times \\ \times & \times & \times & \times & \times & \times & \times \\ \times & \times & \times & \times & \times & \times & \times \\ \times & \times & \times & \times & \times & \times & \times \\ \times & \times & \times & \times & \times & \times & \times \end{pmatrix} \begin{pmatrix} f_0 \\ f_1 \\ \times \\ \times \\ \times \\ f_{N-2} \\ f_{N-1} \end{pmatrix} = \begin{pmatrix} 0 \\ 0 \\ \times \\ \times \\ \times \\ 0 \\ -J_{N-1,N}\, f_N \end{pmatrix} \tag{33}$$

where the $N\times N$ symmetric matrix on the left side is $\left(H_0 + V - E\,\Omega\right)$ and $J_{n,m}(E) = \left(H_0 - E\,\Omega\right)_{n,m}$. We multiply both sides of Eq. (33) by the inverse of this matrix which is the finite $N\times N$ Green's functional matrix $\left(H - E\,\Omega\right)^{-1}$ (i.e., the resolvent of *H*). The last row (row $N-1$) of the resulting matrix equation gives a special relation that determines $\delta(E)$ by using $f_N(E)$ and $f_{N-1}(E)$ from Eq. (32) and reads

$$e^{2\mathrm{i}\delta(E)} = T_{N-1}(E)\,\frac{1 + G_{N-1,N-1}(E) J_{N-1,N}(E) R_N^-(E)}{1 + G_{N-1,N-1}(E) J_{N-1,N}(E) R_N^+(E)} \tag{34}$$

where $T_n(E) = \dfrac{c_n(E) - \mathrm{i}s_n(E)}{c_n(E) + \mathrm{i}s_n(E)}$ and $R_n^\pm(E) = \dfrac{c_n(E) \pm \mathrm{i}s_n(E)}{c_{n-1}(E) \pm \mathrm{i}s_{n-1}(E)}$. To calculate these coefficients, we start with $T_0(E)$ and $R_1^\pm(E)$ then use the recursion relation (31) to obtain the rest†. Since the basis set $\{\psi_n\}$ with the elements given by Eq. (27) are not orthogonal, then the finite Green's function $G_{N-1,N-1}^N(E)$ are obtained by using either formula (11) or formula (17). Thus, we can write

† Using Eq. (A26) in the Appendix of Ref. [3], we can write $T_0(E) = e^{2\mathrm{i}\theta}\,\dfrac{\ell+1+\mathrm{i}t}{\ell+1-\mathrm{i}t}\,\dfrac{{}_2F_1\left(-\ell-\mathrm{i}t,1;\ell+2-\mathrm{i}t;e^{2\mathrm{i}\theta}\right)}{{}_2F_1\left(-\ell+\mathrm{i}t,1;\ell+2+\mathrm{i}t;e^{-2\mathrm{i}\theta}\right)}$ and $R_1^+(E) = \dfrac{e^{-\mathrm{i}\theta}\sqrt{2\ell+2}}{\ell+2+\mathrm{i}t}\,\dfrac{{}_2F_1\left(-\ell+\mathrm{i}t,2;\ell+3+\mathrm{i}t;e^{-2\mathrm{i}\theta}\right)}{{}_2F_1\left(-\ell+\mathrm{i}t,1;\ell+2+\mathrm{i}t;e^{-2\mathrm{i}\theta}\right)}$, where $t = Z/\sqrt{2E}$ and $\cos\theta = \dfrac{8E-\lambda^2}{8E+\lambda^2}$ with $0 < \theta \le \pi$. However, we had to take care of the normalization of the basis element adopted in [3] that differs from ours.

$$G_{N-1,N-1}^{N}(E)=\sum_{j=0}^{N-1}\frac{1}{\sigma_j}\frac{\Gamma_{N-1,j}^2}{\varepsilon_j-E}=\frac{\prod_{i=0}^{N-2}\tilde{\tau}_i\left(\tilde{\varepsilon}_i-E\right)}{\prod_{j=0}^{N-1}\tau_j\left(\varepsilon_j-E\right)}, \tag{35}$$

where $\tilde{\tau}_i=\tau_i^{(N-1,N-1)}$ and $\tilde{\varepsilon}_i=\varepsilon_i^{(N-1,N-1)}$.

Figure 1 is a plot of $|1-S(E)|$, where $S(E)$ is the scattering matrix $e^{2i\delta(E)}$ associated with the short-range potential $V(r)=7.5r^2e^{-r}$ for $\ell=0$ and $Z=0$. We took the computational parameters $\lambda=1.0$ and $N=60$. The figure clearly shows a resonance activity near $E=3.4$. Figure 2 is a plot of the real part (solid curve) and imaginary part (dashed curve) of $S(E)$ for the same system. Figure 3 is a zoom plot of Figure 2 showing that the imaginary part peaks at $E=3.425$, which agrees well with the findings in the literature [4-8] where $E_{res}=3.426-$ $\mathrm{i}0.0128$.

Figure 4 is a plot of $|1-S(E)|$ associated with the short-range potential $V(r)=5e^{-(r-3.5)^2/4}-$ $8e^{-r^2/5}$ for $\ell=0$ and $Z=0$. We took the computational parameters $\lambda=20$ and $N=20$. The figure clearly shows resonance activities near $E=2.2$ and $E=4.6$. Figure 5, which is a zoom plot of the real and imaginary parts of $S(E)$ with $N=100$, gives a better resolution showing that these resonances occur at $E=2.2524$ and $E=4.51$. These are in good agreement with the findings in the literature [8-10] where $E_{res}=2.2524-\mathrm{i}0.00005913$ and $E_{res}=4.5010-$ $\mathrm{i}0.12398$. Figure 6 shows six zoom plots of $S(E)$ for the same system with $\ell=0,1,2$ and $Z=\pm1$. The six resonance energies are in good agreement with the findings in Ref. [10] as shown in Table 1.

Finding the bound states energies, on the other hand, is much easier than finding resonances and does not require any significant increase in the calculation cost. For example, we can choose any desired $(n,m)$ component of the Green's function and just plot $\left|G_{n,m}^{N}(E)\right|$ for $E<0$. The plot will show divergence at the bound state energies even for small $N$ and increasing $N$ will increase the accuracy. Figure 7 is such a plot for $\left|G_{N-1,N-1}^{N}(E)\right|$ associated with the potential $V(r)=5e^{-(r-3.5)^2/4}-8e^{-r^2/5}$ for $\ell=0$, $Z=0$, and $N=5$. The plot shows bound state energies at $E=-4.6$ and $E=-0.8$. Increasing the basis size to $N=15$ and zooming into the figure, we obtain $E=-4.571182831$ and $E=-0.8842806$ which agrees very well with the results in [8] and [10].

Finally, we calculate the energy density of states $\rho(E)$ using its definition as the discontinuity of the $(0,0)$ component of the Green's function across the real axis in the complex energy plane. That is,

$$\rho(E)=\frac{1}{2\pi\mathrm{i}}\lim_{\varepsilon\to0}\left[G_{0,0}(E+\mathrm{i}\varepsilon)-G_{0,0}(E-\mathrm{i}\varepsilon)\right]=\frac{1}{\pi}\mathrm{Im}\,G_{0,0}(E). \tag{36}$$

Several techniques have been developed to obtain a good approximation of $\rho(E)$ using this definition for the finite Green's function $G_{0,0}^{N}(E)$. See, for example, the three techniques outlined in Ref. [11]. For this calculation, we use an alternative orthonormal basis set with the elements

$$\psi_n(r)=\sqrt{\frac{2\lambda\Gamma(n+1)}{\Gamma\left(n+\ell+\frac{3}{2}\right)}}(\lambda r)^{\ell+1}e^{-\lambda^2r^2/2}L_n^{\ell+\frac{1}{2}}(\lambda^2r^2)\,. \tag{37}$$

In this orthonormal basis, the finite Green's function is written using either formula (12) or formula (19) as follows

$$G_{0,0}^N(E)=\sum_{i=0}^{N-1}\frac{\Gamma_{0,i}^2}{\varepsilon_i-E}=\frac{\prod_{i=0}^{N-2}\varepsilon_i^{(0,0)}-E}{\prod_{j=0}^{N-1}\varepsilon_j-E}\,. \tag{38}$$

Figure 8 shows $\rho(E)$ for the system corresponding to the potential $V(r)=7.5r^2e^{-r}$ with $Z=0$ and $\ell=0,1,2$ (left to right) obtained by employing the analytic continuation procedure. In that procedure, $G_{0,0}^N(z)$ is evaluated using (38) in the upper half of the complex energy plane far enough from the discontinuity where it is fitted to a complex analytic function $F(z)$. Subsequently, the function $F(z)$ is analytically continued to the real energy axis to replace $G_{0,0}(z)$ in (36) for evaluating $\rho(E)$.

In closing, we add a couple of technical notes. The first is about calculating the matrix elements of the potential function, $V_{n,m}=\langle\psi_n|V|\psi_m\rangle$, which is needed for the evaluation of the total Hamiltonian matrix $H_{n,m}=\left(H_0\right)_{n,m}+V_{n,m}$. We have used Gauss quadrature integral approximation associated with the Laguerre polynomials (see, for example, Section 2 of Ref. [12] for details of this technique). The second note is about the calculation of deep resonances (those with large negative imaginary parts). One could use complex scaling/rotation where the scale parameter $\lambda$ is replaced by $\lambda e^{-i\varphi}$ with $0<\varphi<\pi/4$.

# 4. Conclusion

In this work, we made a detailed and complete calculation of the finite matrix representation of the resolvent operator in any square-integrable basis. The work is useful for scientist and engineers who are looking for such treatment in one place and whether the basis chosen is orthogonal or not. Formulas (17) and (21) are particularly valuable because of the associated reduction in computational cost. An interesting byproduct of this study is an expression for the elements of the normalized eigenvectors (generalized eigenvectors) in terms of its eigenvalues (generalized eigenvalues) as given by (23) and (25), respectively.

As a physical application, we showed how to use these findings is calculating the resonance structure, bound state energies, and density of states in quantum mechanics.

# Acknowledgement

I am grateful to Prof. H. A. Yamani for assistance in obtaining formula (21) for the finite Green's function $G_{n,m}^N(z)$ in an orthonormal basis with $n\neq m$.

# References

[1] D. M. Mitnik and S. A. H. Mitnik, *Wavefunctions from energies: Applications in simple potentials*, J. Math. Phys. **61**, 062101 (2020) and references therein

[2] A. D. Alhaidari, E. J. Heller, H. A. Yamani, and M. S. Abdelmonem (Eds.), *The J-Matrix Method: Developments and Applications* (Springer, Dordrecht, Netherlands, 2008)

[3] H. A. Yamani and L. Fishman, *J-matrix Method: Extension to Arbitrary Angular Momentum and to Coulomb Scattering*, J. Math. Phys. **16** (1975) 410

[4] A. D. Isaacson, C. M. McCurdy and W. H. Miller, Chem. Phys. **34** (1978) 311

[5] C. H. Maier, L. S. Cederbaum and W. Domcke, *A spherical-box approach to resonances*, J. Phys. B: At. Mol. Phys. **13** (1980) L119

[6] V. A. Mandelshtam, T. R. Ravuri and H. S. Taylor, *Calculation of the density of resonance states using the stabilization method*, Phys. Rev. Lett. **70** (1993) 1932

[7] H. A. Yamani and M. S. Abdelmonem, *Characterization of resonances using an exact model S-matrix*, J. Phys. A: Math. Gen. **28** (1995) 2709

[8] S. A. Sofianos and S. A. Rakityansky, *Exact method for locating potential resonances and Regge trajectories*, J. Phys. A: Math. Gen. **30** (1997) 3725

[9] C. H. Maier, L. S. Cederbaum and W. Domcke, *A spherical-box approach to resonances*, J. Phys. B: At. Mol. Phys. **13** (1980) L119

[10] A. D. Alhaidari, *Unified algebraic treatment of resonance*, Int. J. Mod. Phys. A **20** (2005) 2657

[11] H. A. Yamani, M. S. Abdelmonem, and A. D. Al-Haidari, *Extracting density information from finite Hamiltonian matrices*, Phys. Rev. A **62** (2000) 052103

[12] A. D. Alhaidari, *Gauss Quadrature for Integrals and Sums*, Int. J. Pure Appl. Math. Res. **3** (2023) 1

## Figures Captions

**Fig. 1**: Plot of $|1-S(E)|$ for the system associated with $V(r)=7.5r^2e^{-r}$ for $\ell=0$ and $Z=0$. We took the computational parameters $\lambda=1$ and $N=60$. The figure shows a resonance activity at $E=3.4$.

**Fig. 2**: A reproduction of Figure 1 but for the real part (solid curve) and imaginary part (dashed curve) of $S(E)$.

**Fig. 3**: Zoom plot of Figure 2 showing that the imaginary part peaks at $E=3.425$, which agrees well with the location of this resonance found elsewhere [3-7]. The horizontal axis is the energy in atomic units.

**Fig. 4**: Plot of $|1-S(E)|$ associated with the short-range potential $V(r)=5e^{-(r-3.5)^2/4}-8e^{-r^2/5}$ for $\ell=0$ and $Z=0$. We took the computational parameters $\lambda=20$ and $N=20$. The figure shows resonance activities at $E=2.2$ and $E=4.6$.

**Fig. 5**: Zoom plots of Figure 4 for the real and imaginary parts of $S(E)$ showing resonance activities at $E=2.2524$ and $E=4.51$. We took the computational parameters $\lambda=20$ and $N=100$. The horizontal axis is the energy in atomic units.

**Fig. 6**: Zoom plots of the real and imaginary parts of $S(E)$ for the system associated with the potential $V(r)=5e^{-(r-3.5)^2/4}-8e^{-r^2/5}$ for $Z=-1$ (top row) and $Z=+1$ (bottom row). The angular momenta were taken as $\ell=0,1,2$ (from left to right columns). The horizontal axis is the energy in atomic units.

**Fig. 7**: Plot of $\left|G^N_{N-1,N-1}(E)\right|$ for the system with $V(r)=5e^{-(r-3.5)^2/4}-8e^{-r^2/5}$ and for $\ell=0$, $Z=0$ and $N=5$. The plot blows up at the two bound state energies: $E=-4.6$ and $E=-0.8$.

**Fig. 8**: Energy density of states $\rho(E)$ for the system corresponding to $V(r)=7.5r^2e^{-r}$ with $Z=0$ and $\ell=0,1,2$ (from left to right). The horizontal axis is the energy $E$ in atomic units.

## Table Caption

**Table 1**: Resonance energies associated with the potential $V(r)=5e^{-(r-3.5)^2/4}-8e^{-r^2/5}$ for the given electric charge and angular momentum. These results are obtained from Figure 6 and compared to those in the Table of Ref. [10]. Uncertainty in the numbers obtained from Figure 6 is in the last decimal digit. The smaller the imaginary part of the resonance energy, the sharper the resonance, and the larger the number of significant digits.

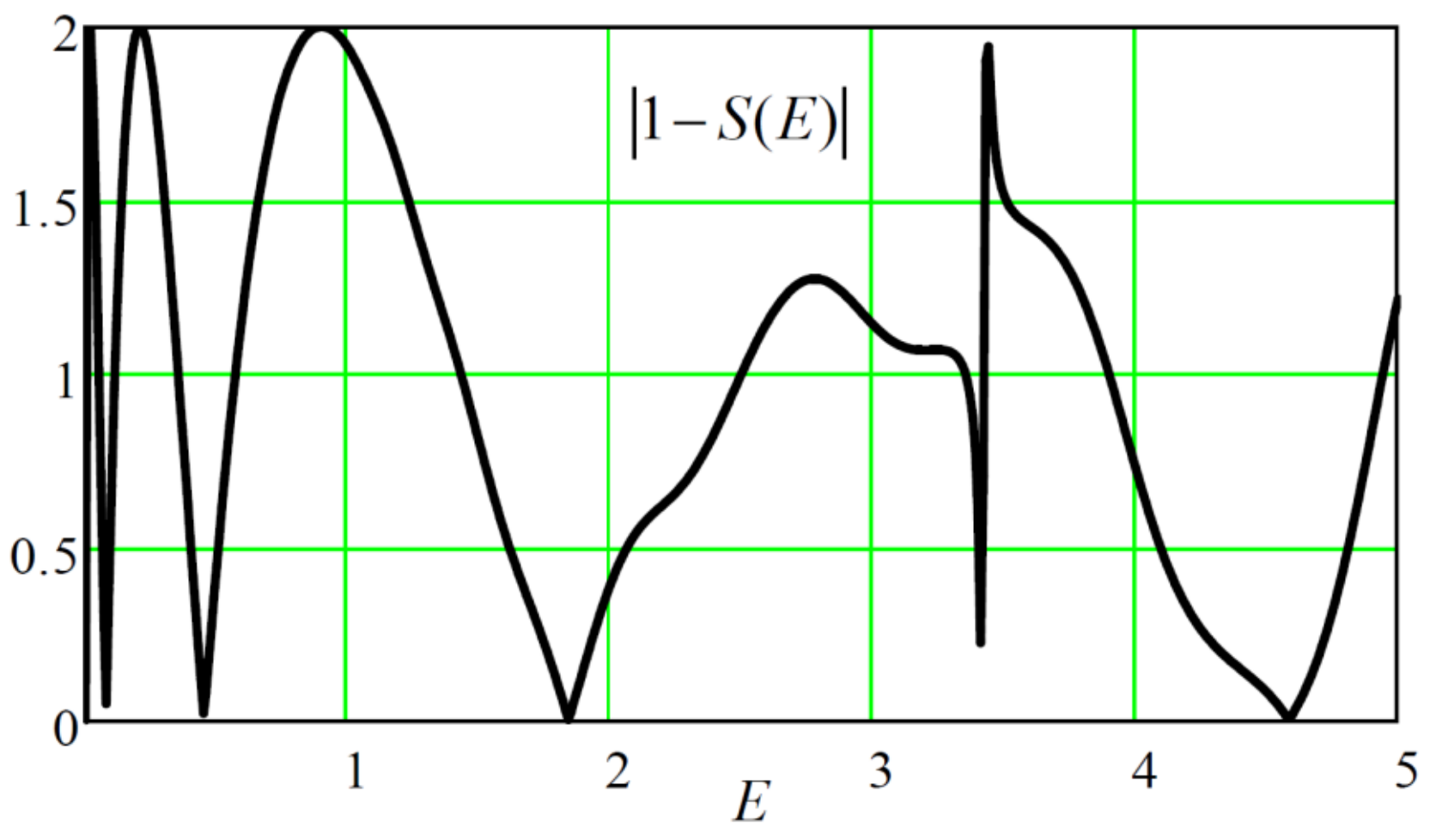

**Fig. 1**

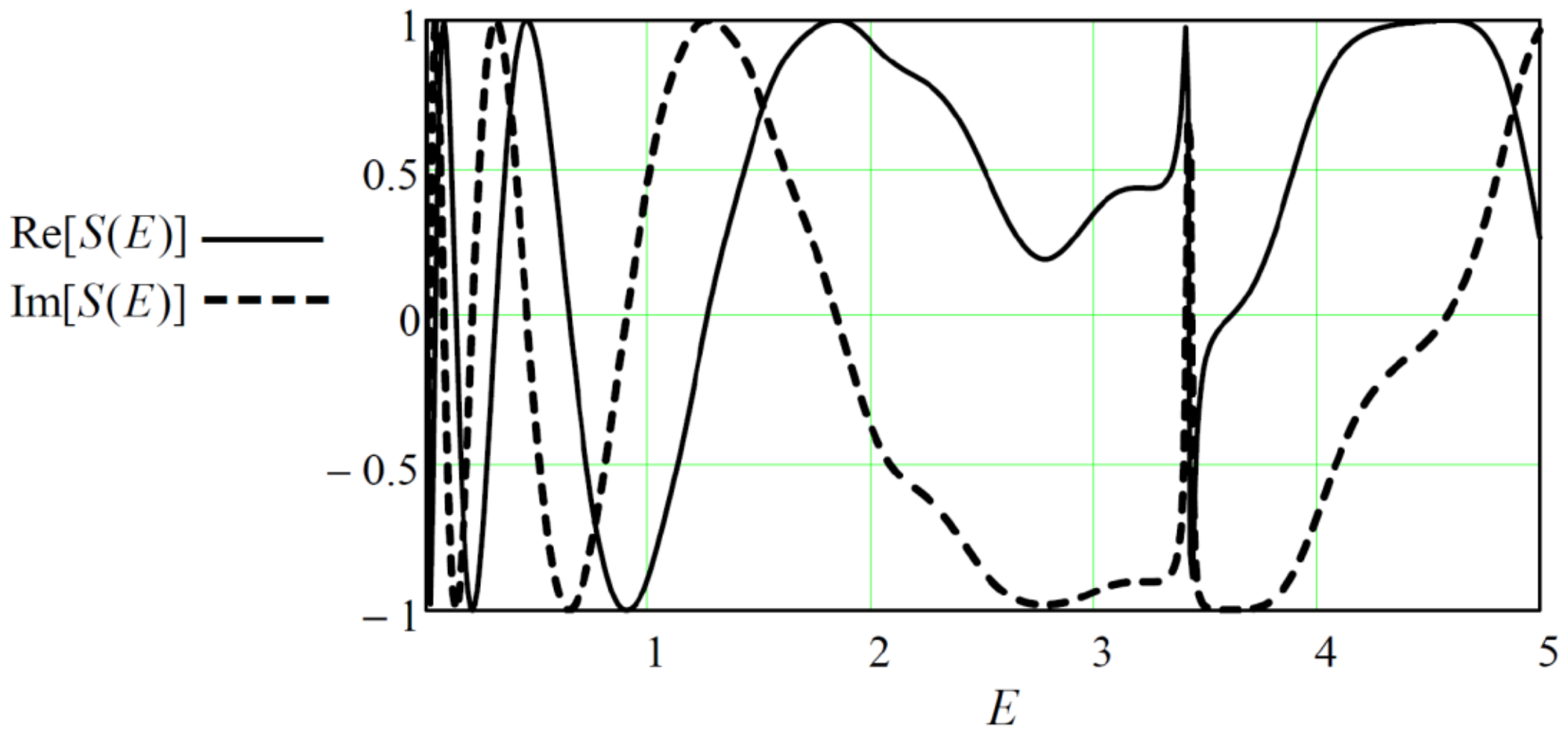

**Fig. 2**

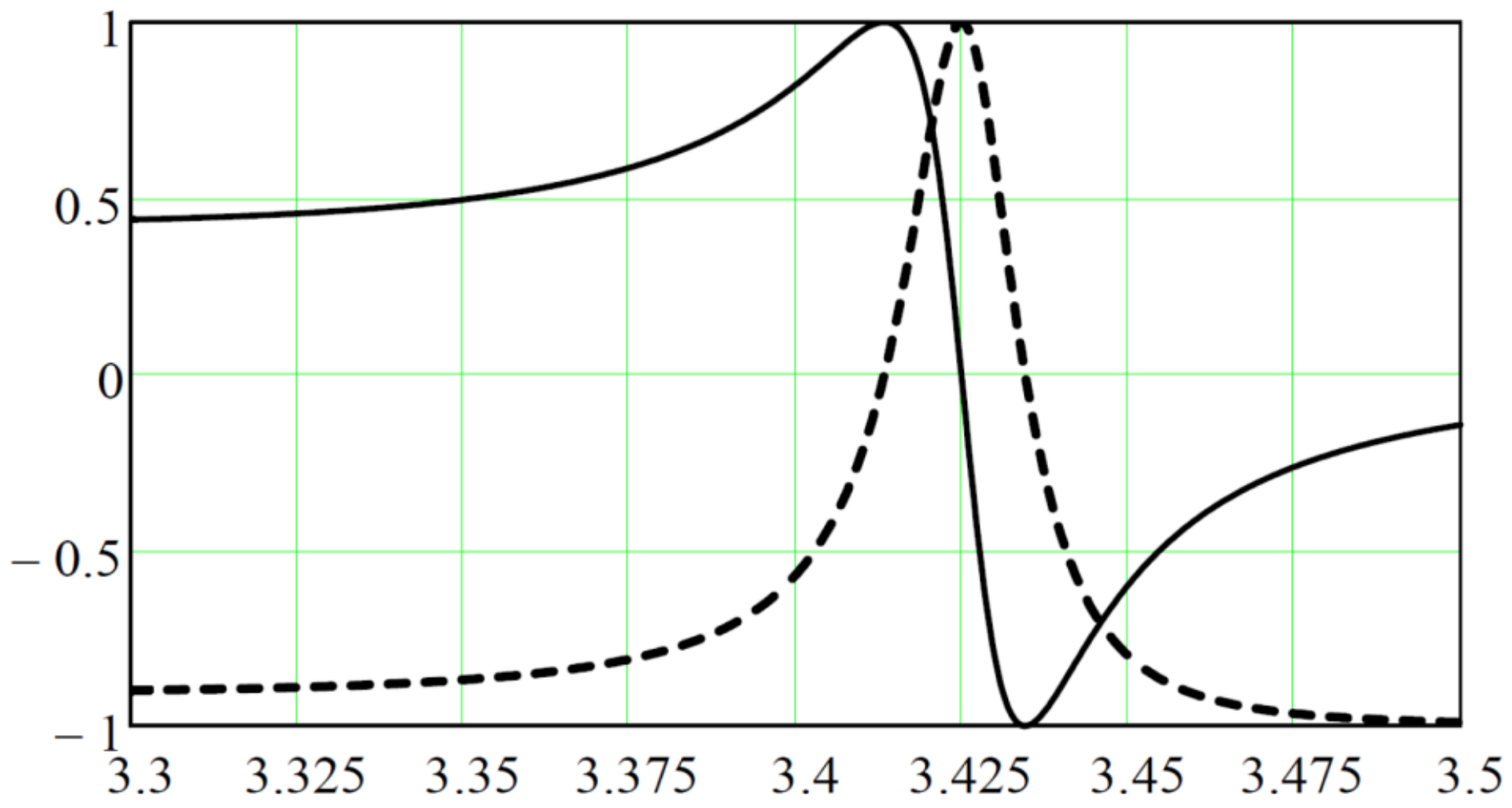

**Fig. 3**

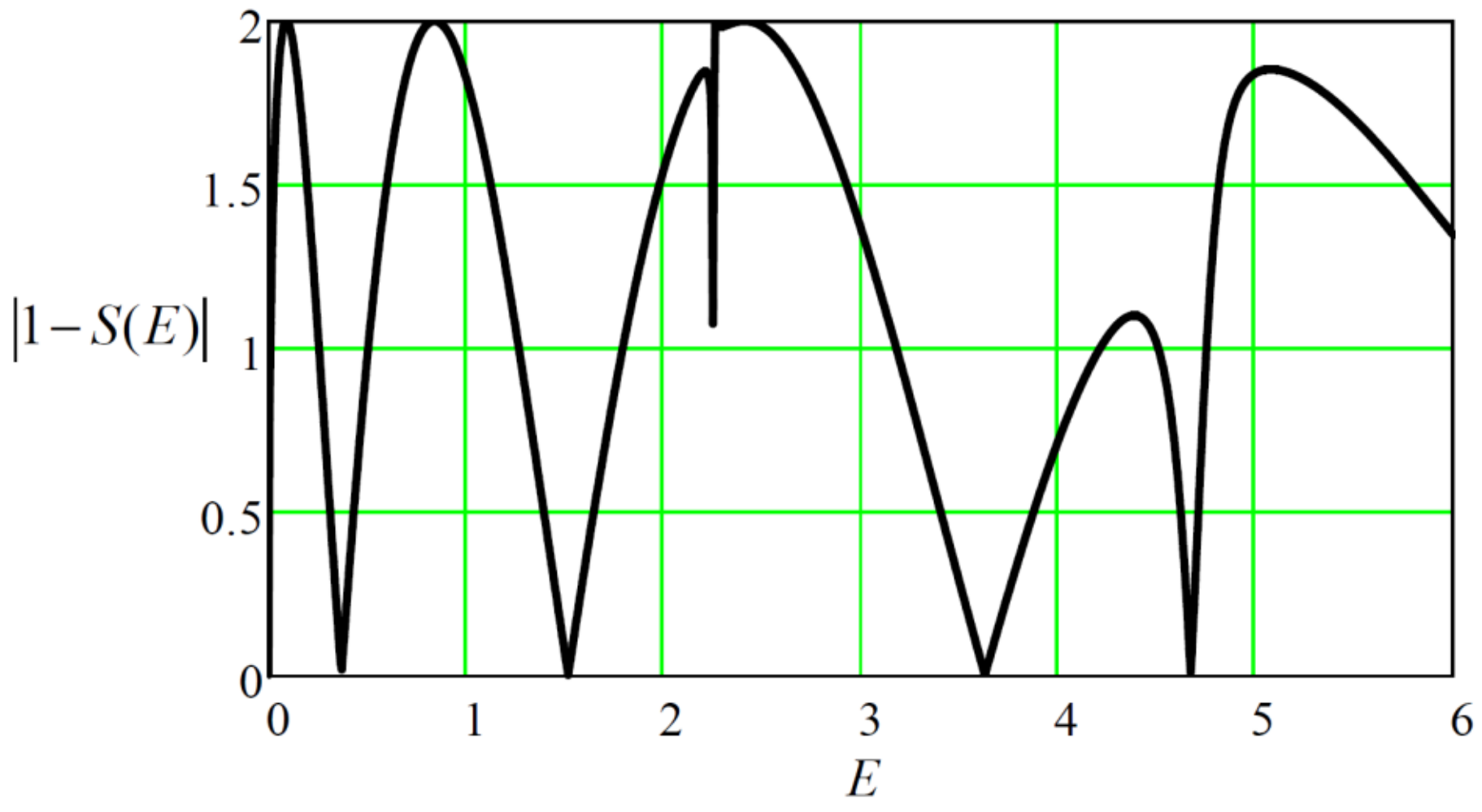

**Fig. 4**

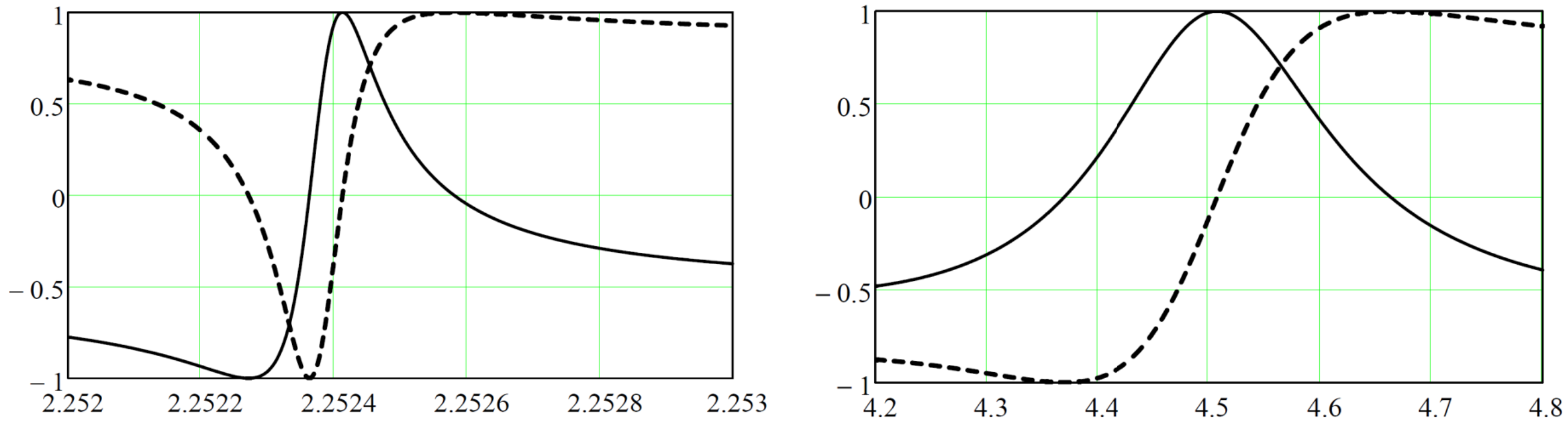

**Fig. 5**

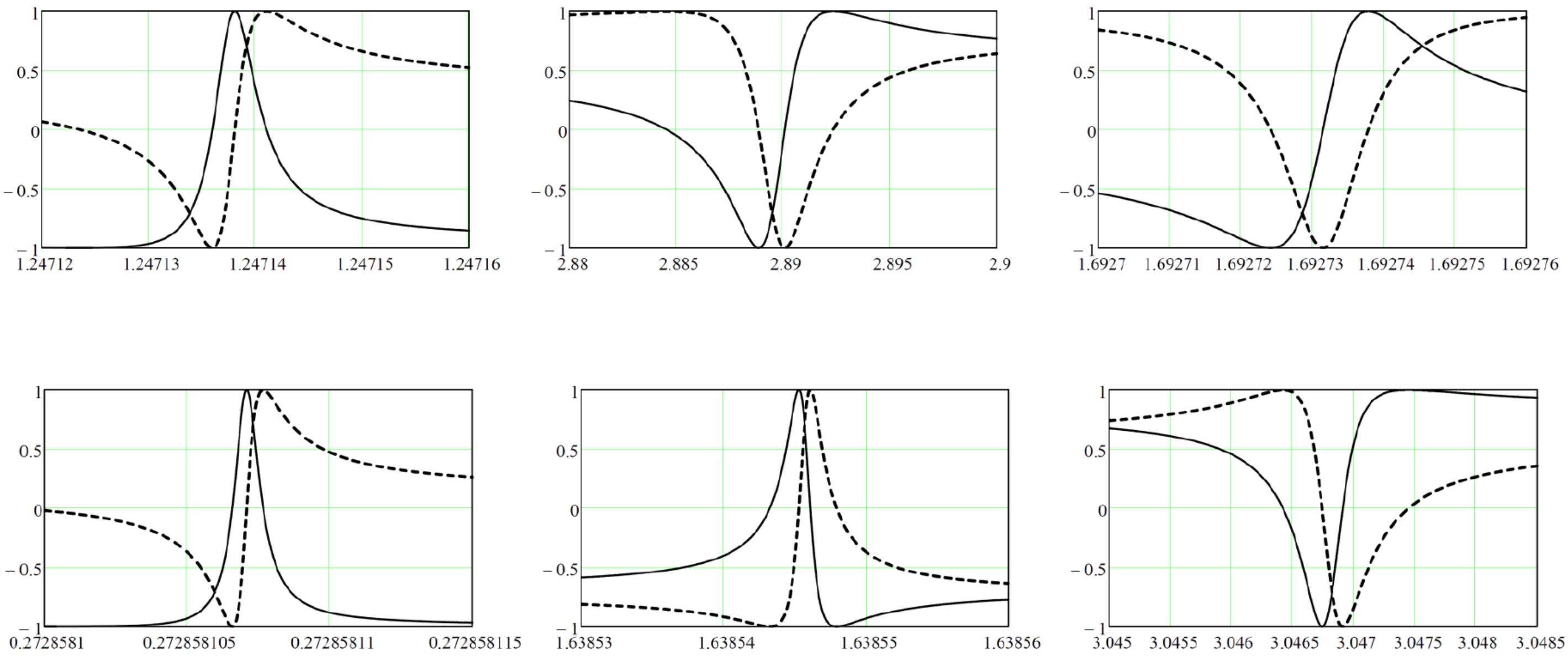

**Fig. 6**

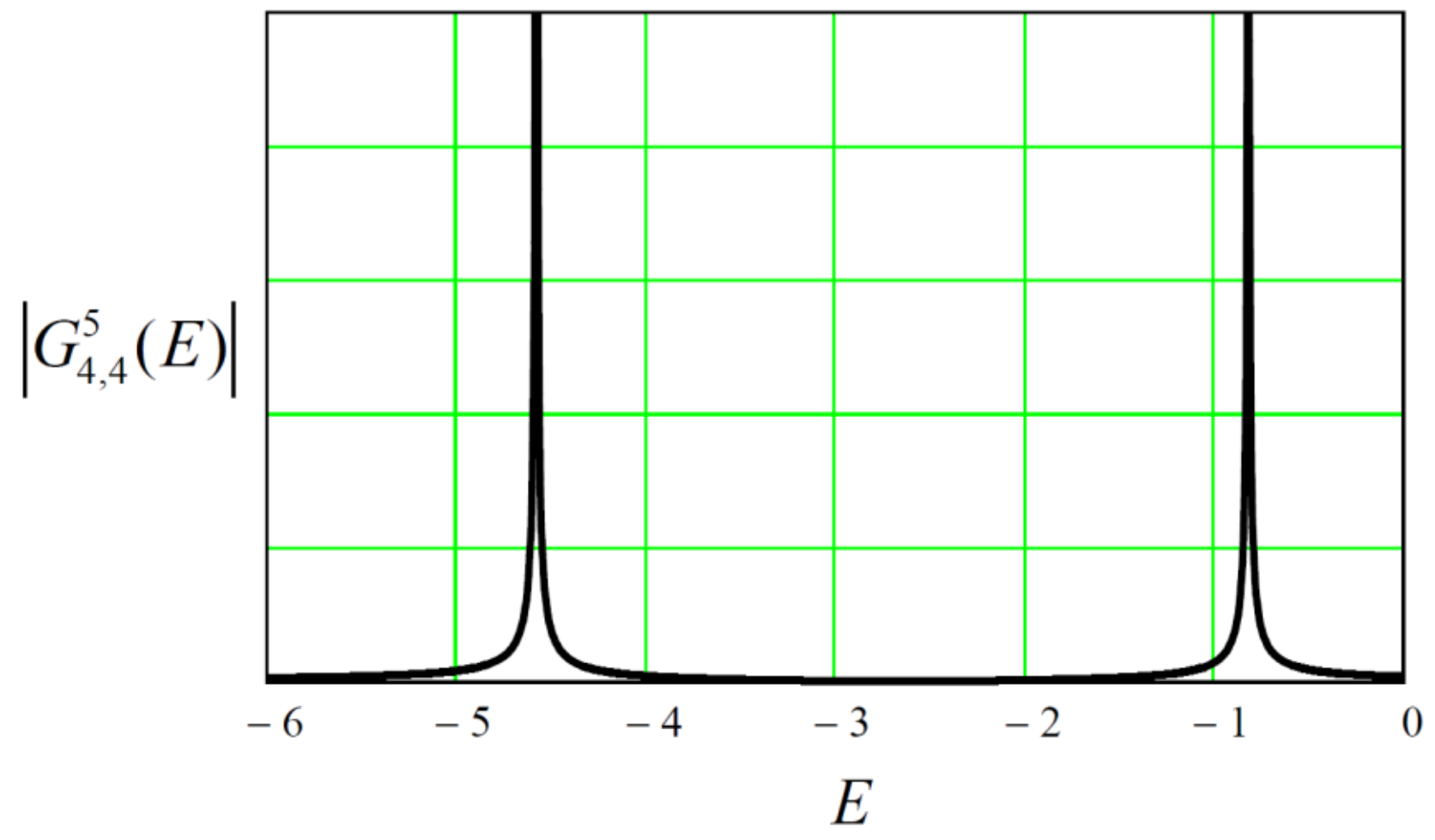

**Fig. 7**

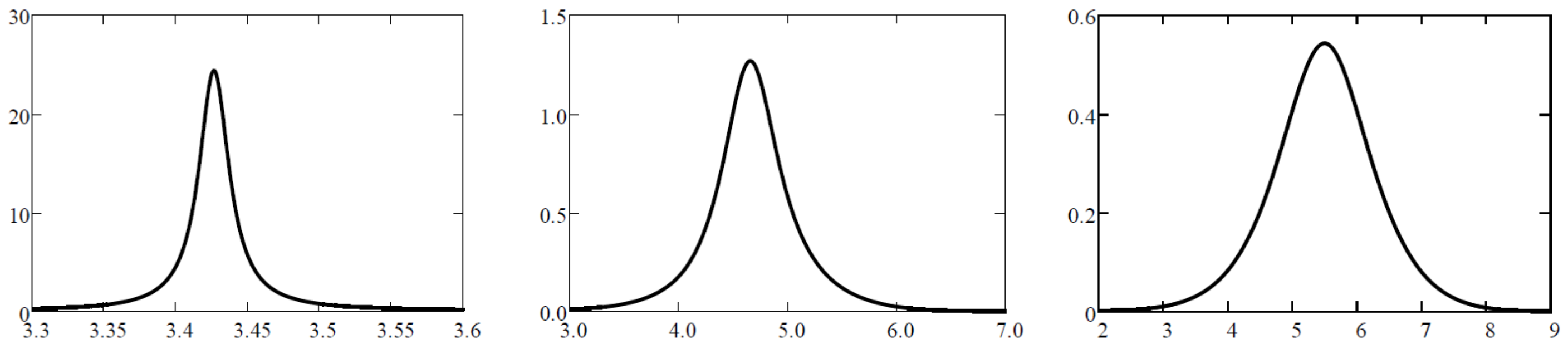

**Fig. 8**

**Table 1**

| $Z$ | $\ell$ | $\mathrm{Re}(E_{res})$ (Fig. 6) | $E_{res}$ (Ref. [10]) |
|---|---|---|---|
| −1 | 0 | 1.247137 | 1.247137679 − i0.0000023985 |
| | 1 | 2.889 | 2.889663069 − i0.0014603245 |
| | 2 | 1.69273 | 1.692732086 − i0.000006761 |
| +1 | 0 | 0.272858107 | 0.272858107 − i0.0000000006 |
| | 1 | 1.638546 | 1.638545711 − i0.000000971 |
| | 2 | 3.0467 | 3.046808400 − i0.000100441 |